# Decentralized Policy Information Points for Multi-Domain Environments

M Ridwanur Rahman
*Faculty of Information Technology*
*Monash University*
Melbourne, Australia
mrah0018@student.monash.edu

Ahmad Salehi S.
*Faculty of Information Technology*
*Monash University*
Melbourne, Australia
ahmad.salehishahraki@monash.edu

Carsten Rudolph
*Faculty of Information Technology*
*Monash University*
Melbourne, Australia
carsten.rudolph@monash.edu

*Abstract*—Access control models have been developed to control authorized access to sensitive resources. This control of access is important as there is now a need for collaborative resource sharing between multiple organizations over open environments like the internet. Although there are multiple access control models that are being widely used, these models are providing access control within a closed environment i.e. within the organization using it. These models have restricted capabilities in providing access control in open environments. Attribute-Based Access Control (ABAC) has emerged as a powerful access control model to bring fine-grained authorization to organizations which possess sensitive data and resources and want to collaborate over open environments. In an ABAC system, access to resources that an organization possess can be controlled by applying policies on attributes of the users. These policies are conditions that need to be satisfied by the requester in order to gain access to the resource. In this paper, we provide an introduction to ABAC and by carrying forward the architecture of ABAC, we propose a Decentralized Policy Information Point (PIP) model. Our model proposes the decentralization of PIP, which is an entity of the ABAC model that allows the storage and query of user-attributes and enforces fine-grained access control for controlling the access of sensitive resources over multiple-domains. Our model makes use of the concept of a cryptographic primitive called Attribute Based Signature (ABS) to keep the identities of the users involved, private. Our model can be used for collaborative resource sharing over the internet. The evaluation of our model is also discussed to reflect the application of the proposed decentralized PIP model.

## I. Introduction

Nowadays, data is everywhere. With the exponential rise of data, the need for collaborative relationships and the sharing of resources has become a necessity as the only way to leverage data is by securely sharing it across open environments like the internet. An open environment means multiple domains with different security and privacy requirements can collaborate and share resources. Data Owners (DO) like Businesses, Industries or Academia who possess data, operate their own administrative domains, called security domains, where they control access to their own data and resources and work independently of other domains but want to collaborate and share resources [1]. This collaboration and sharing of resources is important as the data that the different domains possess can be used for new findings or to continue studies [2].

DOs have been outsourcing their data to the cloud because of its ease of use and reliability. However, the cloud works on the idea that the parties that want to collaborate, are in the same security domain. This creates limitations in access control (AC) which is a major security vulnerability. The AC models that are in use like DAC (Discretionary Access Control), MAC (Mandatory Access Control) and RBAC (Role Based Access Control) are centralized and focus on the protection of data in an environment where the details of the user are already stored in the system. Their main limitation is that they have drawbacks in open environments like the internet [3].

A solution to the problem of AC in open environments is to exchange attributes. ABAC was introduced with the idea that access control can be established based on present attributes of object, subject, action and environment of users [4].

In an ABAC system, the attributes of the users are queried from the database by the Policy Information Point (PIP). In a multi domain environment, different domains have different security configurations [5]. However, the domains need to have a system to query their attributes to be granted permissions to access resources in other domains. This is the main advantage of a decentralized PIP. Domains would have their own security configurations but the feature of querying the attribute remains the same, as provided by the PIP. Also if implemented, the architecture of ABAC would ensure the same entities work together in the domains. This ensures consistency and makes the implementation easy. However, privacy is required otherwise users can be identified based on their attributes. To keep the attributes secure, cryptographic primitives must be applied. Attribute Based Signature (ABS) was proposed in 2010 by Maji, Prabhakaran and Rosulek [6]. ABS is a cryptographic primitive where, a user, who possesses attributes, can sign a message using a key based on the predicate that is satisfied by their attributes. This signature only reveals that a single user who possess a set of attributes that satisfy a predicate has attested to the message. The signature keeps the attributes of the user secure. So, the identity of the user stays unknown. Therefore, we believe that an ABS scheme can be added to an ABAC model to keep the privacy of the users between multiple domains. This paper is not concerned with the cryptographic processes involved in signature generation.

Which is why we did not include mathematical equations required to generate signatures. We assume that the appropriate signature generation and verification process works correctly. For this reason, we used an open source implementation of ABS [7].

While many works have explored the applications of ABAC for existing AC problems [8], [9], few have looked into the secure transfer of the attributes in multiple domains [10]. In this paper, we propose a general approach to the AC of resources by the enforcement of ABAC and apply the concept of ABS to implement security for the transfer of attributes between domains. The main advantage of this model is the decentralization of PIP which allows the ABAC model to run within different security configurations as set by the domain. The fundamental challenge that each domain who want to run their own PIP, thus making a decentralized network of PIPs, is to make their PIPs accessible to other domains and to make them efficient and functional to be used with ease. The goal of this research is to show proof of a fine-grained AC model that will extend the functionalities of secure resource sharing in open environments where the details of the users are not known. We entitle the main contributions of this paper as follows:

- We propose an idea for a Distributed PIP which stores the attributes of the users of the domain.
- We carry forward the architecture of ABAC and focus on the development of a Decentralized Policy Information Point for ABAC. This can be used to provide fine-grained authorization to share data over the internet as it makes use of attributes.
- We use the concept of ABS as a form of cryptographic scheme to ensure both identity-anonymity and attribute-anonymity which would guarantee privacy for the user.

The rest of the paper is organised as follows. Section II talks about related work done in the field of ABAC and ABS. Section III provides the background of the access control models that are in use. Section IV defines our proposed Decentralized PIP model. Section V shows the configurations and frameworks we used to implement our model. Section VI shows the evaluation of our model. In Section VII, we conclude the paper with a summary and future work.

## II. RELATED WORK

This section talks about the research works that used ABAC and ABS. Some of the works provide different levels of fine-grained authorization. However, none of the implementations provide a decentralized PIP. Our model is the first to propose a decentralized PIP

### A. ABAC in open environments

The concept of ABAC has been found to be suitable for open environments like the internet where resource sharing is important and AC, based on identity, is not enough [11]. This is where the use of attributes are more suitable [12]. In recent years, a lot of work has been conducted on ABAC [8], [9], [13]. These works list the deficiencies of the traditional access control models for web services like MAC, DAC and RBAC as insufficient in dynamicity, unscalable in regards to the growing business and limited to single domains. RBAC suffers from 'role-explosion' and 'role-expansion'. This is a highly undesirable situation where the growth of a company increases the roles exponentially which hampers their system. In this situation, the system becomes overloaded and in the worst case, might need to be changed which is expensive[14], [15], [16]. Most of these works cite ABAC as being the appropriate model for open environments because it is semantically expressive, scalable and flexible [17].

Seol, K. el al., implemented ABAC for sharing sensitive Electronic Health Records (EHR) between hospitals in the cloud [8]. EHRs are designed to allow interoperability between multiple hospitals, laboratories, pharmacies and universities and are very important to share [18]. However, since EHR data contain sensitive and confidential information, their access needs to be controlled so that only authorized doctors can access them. The authors mentioned the inefficiency of RBAC as being inflexible and proposed an ABAC model for AC, where the security policies were written in XACML. In their model, they used an open source ABAC model developed by WSO2 and added cryptographic schemes like XML encryption and XML digital signature for encrypting their data so it cannot be seen by unwanted users. WSO2 is a company that has built commercial and open source ABAC systems.

Although the authors' work is interesting as they provide fine-grained access control and security over multi-domains, it is unclear what attributes the first XACML request would contain. Different domains would contain numerous resources which need various attributes. Some might need a combination of ten attribute while some resource which are not very important can have 2 attributes in its policy. This is a clear gap in an ABAC system as not knowing the number of attributes to keep in the first request, one domain might send more or fewer attributes than is needed. This would add an overhead as sending too many attributes slow down the web service while sending fewer attributes would result in deny response. Which would increase the number of requests-responses. The privacy issue of exchanging attributes between domains was not mentioned in their work.

### B. Applications of ABS in multi-domains

There are signature schemes like Group Signatures, Ring Signatures [6]. Group signature require a group manager to create groups. The main limitation is if an organization is using group signatures, some members might not have been added. Ring signatures allow the user to choose any other signer including himself, and sign any message either by using his own secret key and the others' public key. This process does not require their approval or assistance so any user can use any other signer which is not secure [19].

ABS has been implemented to provide security for distributed open environments like blockchain and IOT, [20], [21]. ABS is a good tool to provide security for our model as it supports multi-authority environments. A few works have been done on similar environment [22], [23]. These works on multi-authority state that having a central authority to manage other authorities from where different secret keys are queried, has a major security issue as the central authority can be compromised. This is the main issue with ABE which makes use of central authority [24].

From these works, it can be clearly seen that ABAC plays a much better role in practice as compared to traditional AC models in decentralized and open environments. This is our reason for choosing ABAC. However, since ABAC works with exchanging attributes, security is an issue. ABS is suitable to eliminate this issue since it keeps the user's identifying information (attributes) secure and only reveals the user who satisfies the AC policy wants to access the resource. Apart from that if ABS is applied, users cannot lie about their attributes as they are signed by their signature key which is generated by an attribute authority. In ABS, a verifier can create a predicate and no attributes are revealed. Even if the signatures do not match, the verifier does not learn the attributes. Hence, attributes remain completely hidden and fine-grained access control is imposed.

## III. BACKGROUND

TABLE I
LIST OF ABBREVIATIONS USED IN THIS PAPER AND THEIR ACRONYMS.

| | |
|---|---|
| ABAC | Attribute Based Access Control |
| ABE | Attribute Based Encryption |
| ABS | Attribute Based Signature |
| AC | Access Control |
| API | Application Programming Interface |
| APK | Attribute Public Key |
| ASK | Attribute Signing Key |
| CP-ABE | Ciphertext-Policy Attribute Based Encryption |
| DAC | Discretionary Access Control |
| DO | Data Owner |
| EHR | Electronic Health Record |
| KP-ABE | Key-Policy Attribute Based Encryption |
| JSON | Javascript Object Notation |
| MAC | Mandatory Access Control |
| NIST | National Institute of Standards and Technology |
| OASIS | Organization for the Advancement of Structured Information Standards |
| PEP | Policy Enforcement Point |
| PDP | Policy Decision Point |
| PAP | Policy Administration Point |
| PIP | Policy Information Point |
| RBAC | Role Based Access Control |
| REST | Representational State Transfer |
| SKA | Signing Key |
| TPK | Trustee Public Key |
| UML | Unified Modeling Language |
| XACML | eXtensible Access Control Markup Language |
| XML | Extensible Markup Language |

This section explores the concept and architecture of ABAC and ABS. A list of abbreviations used in this paper and their acronyms are depicted in Table 1.

### A. Attribute-Based Access Control (ABAC)

According to the National Institute of Standards and Technology (NIST), ABAC is an access control method where access permissions to resources are granted or denied by policies which are based on assigned attributes containing values of the subject, object, environment conditions and the names of resources [25]. Since AC is a major limitation in web applications that cater to open environments, we will be applying ABAC on the authorization layer of a web application which possess sensitive resources. This means, let's say an API (Application Programming Interface) of a web application has sensitive data. We want to share these data to authorized users over open environments. The administrator of the application will write a policy to protect the resource. For authorization, the requester would need to provide his attributes. The attributes are then sent to the ABAC system which decides whether to permit or deny the access request by checking if the attributes of the requester are reflected on the policy of the resource.

The access request has to contain attributes. An attribute is a statement about a user. For example, an attribute can be name, age, gender (male or female), state of residence of an individual [26]. These attributes need to be contained in a specific format based on categories. The attributes can be classified into the categories: Subject, Action, Resource and Environment. Subject category can hold attributes like username of the user, role, email address and age. Action category can hold what action the user wants to perform on the resource which can be create, read, update or delete. Resource category can hold the name of the resource they want to access, name of folder or path of the file. Environment category can hold the time of day or location.

Access policies or rules can be written using these attribute values. In an access policy, the administrator would set up different attributes based on the categories. These policies are usually written using a policy language. Out of which XACML (eXtensible Access Control Markup Language) is the best known and is mentioned in the NIST publication on ABAC [27], [25].

XACML is an XML-based language for access control that has been standardized by the Technical Committee of the OASIS consortium [28]. It describes the access control policy. XACML has a rich data typing model and allows stating complex conditions.

An ABAC system constitutes of multiple domain entities. Each domain entity performs a specific function. They are presented in Figure 1.

The process of ABAC (Figure 1) is explained as follows:
1) A request comes to the PEP via RESTful API to view a resource. For the case of simplicity, this request contains attributes and wants to access a specific resource. PEP intercepts the request and sends it to the context handler.
2) The context handler forwards the request to the PDP.
3) The PDP receives the request. Sometimes, it would not have enough attributes needed to make the decision. In which case, it asks the PIP for the missing attributes.
4) Since the PIP is responsible for storing the attributes, it receives an attribute query request from the PDP. It searches the local database and responds with the attributes.
5) The PDP, with all the attributes now present, will check the policy stored in the PAP for that specific resource. It then will check if the attributes are reflected in the policies of the resources. Then it will generate an access/deny response which it passes to the PEP. [29].
6) Context Hanlder forwards the decision to the PEP.
7) The PEP checks the response. If Accept or True (depending upon how the ABS sends a positive response), it lets the user access the resource; otherwise, it denies the request (Deny/False).

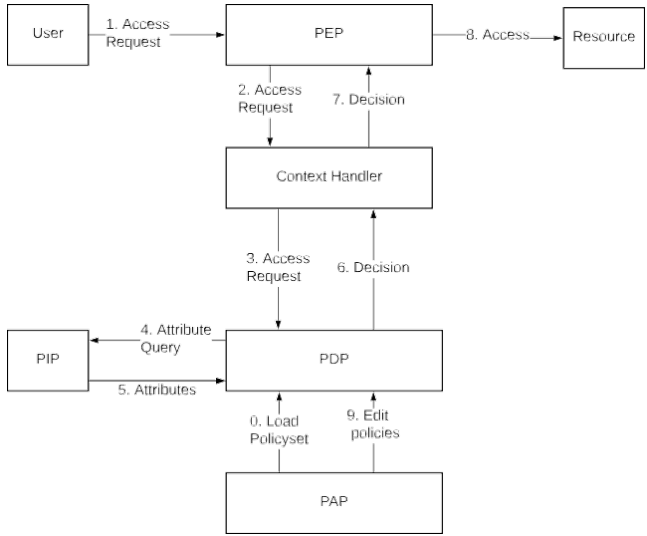

Fig. 1. ABAC Architecture as standardised by NIST [25].

### B. Attribute-Based Signature (ABS)

ABS hides the attributes and all other information that can identify the user. It is suitable to implement ABS in applications that require both data authentication and the preservation of privacy [21]. To generate ABS, as explained in [6]:

- ABS.TSetup: This is run by the signature trustee. It generates a trustee public key (TPK).
- ABS.ASetup: This is run by the attribute authority. It generates a key pair, attribute public key (APK) and attribute signing key (ASK).
- ABS.AttrGen: Takes the ASK as input and the attributes, this will generate the signing key (SKA).
- ABS.Sign: Takes the TPK, APK, SKA, message and predicate as input. A predicate is a combination of all the attributes. For simplicity, in our prototype, we used only the conjunction of the attributes. An ABS scheme in this process, by taking TPK, APK, SKA, message and the predicate as input, outputs a signature.
- Verify: Takes the TPK, APK, message and predicate as input. Outputs the value True/False.

## IV. PROPOSED MODEL

This section talks about our proposed Decentralised PIP model and explains the fundamental concepts behind it.

### A. System Model

In this section, we describe how our decentralized PIP model works. A scenario for our work would be: multiple domains exist possessing resources. Each domain will be able to view the names of the resources that the other domains possess. However, since some resources are sensitive, the administrator of the domain can set access control policiesto control the access of those resources. These policies are conditions or rules. These conditions need to be matched by the attributes of the users of other domains in order for themto gain access. Although our proposed model can be applied inmulti domains, for conveying the functionality of the model, our scenario is based on a peer-to-peer environment where thedomains both know each other.

Requester from domain 2 wants to view a resource in domain 1. It sends a request. Domain 1 checks the policyof the resource and responds with the names of the attributes needed in the form of a predicate statement. Domain 2 loads the attributes, generates a claim-predicate that satisfies the predicate, then signs a message and sends it. Domain 1 verifies the signature and either allows or rejects the access to the request. This process is depicted in Figure 2.

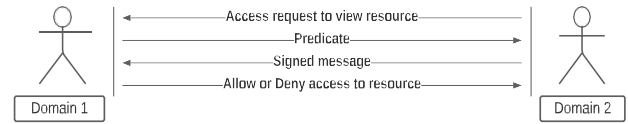

Fig. 2. Interaction between two domains in the proposed Decentralized PIP model.

### B. Proposed Methodology

The fundamental steps of our proposed model are listed here:
- Administrator will write policies for resources. This policy will be written in a name-value pair.
- Requester from a different domain, who wants to view the resource, will get a predicate which would contain the names of attributes needed to view the resource. This predicate is generated from the attribute names in the policy of that resource as set by the administrator.
- Requester would collect the attributes needed to satisfy the predicate from their PIP. This essentially forms the base of our Decentralized PIP model. The PIPs are used to query attributes. In a multi-domain environment, differ-ent domains reside. So for collaboration, these domains would query their PIPs for attributes. All these PIPs are not controlled by a single domain rather they arecontrolled by their own domain.

Using the attributes that are returned from the PIP, the requester would generate their claim-predicate statement.

- Then sign a message with the claim-predicate statement. This signing is accomplished with the open source ABS.
- The message will be sent to the domain that contains the resource using HTTP methods.
- The domain will verify the signature and permit or deny the access to the resource. This is also accomplished with the open source ABS.

*1) Predicate Management:* Bob is an administrator of domain 1. He enters a new resource called 'Resource 1'. Bob has the option to either make it accessible for all domains or he can protect it by writing a policy. In a policy, Bob would have to enter attributes according to category Subject, Action, Resource and Environment. The Resource category will hold the identifier of the resource which could be the name/unique id/file path. For each attribute, he would need to enter the name of the attribute as well as the attribute value. This means, each resource will have a name-value pair of attributes where the name is the attribute name and the value is the attribute value. Lets say resource 1 can only be read by Alice who is a cardiologist from the Box Hill hospital in Melbourne. Alice's subject will be Alice, cardiologist and Box Hill hospital. Her environment will be Melbourne. So, to enter this policy, Bob would need to enter names of each of the attributes. This means, attribute value Alice will have the attribute name 'first name', cardiologist attribute value will have the attribute name 'position' or 'doctor type', 'Box Hill' attribute name would have the attribute value 'hospital' and Melbourne attribute name will have the attribute value 'city'. The structure is shown here:

```
first_name: Alice
position: cardiologist
hospital: Box Hill
city: Melbourne
```

Different domains would design their databases differently where the variable names for attributes might be different. For example, the database field 'first_name' in one domain might be named 'fname' in another domain. This is why the name-value pair of the attribute is important as this would keep the naming consistent. Although the name-value pair is useful, there needs to be a mapping process for the names in other domains.

*2) Accessing Resource:* With the predicate statement provided from domain 1, Alice's PEP will query her PIP to check if the required attributes exist or not. For our scenario, we are assuming she has the specific attributes required to satisfy the predicate and the PIP returns the specific attributes to the PEP. Then, using the set of attributes, her PEP will generate a claim-predicate statement which satisfy the pred- icate statement from domain 1. An example of her claim- predicate statement is: ALICE AND CARDIOLOGIST AND BOX HILL AND MELBOURNE. Using this claim-predicate, her PEP will sign a message. This message endorses that Alice has the specific attributes needed to satisfy the predicate from domain 1. This message will be serialized and transferred to domain 1. Domain 1's PEP will verify the signature based on the attribute values set on the policy. The verifier outputs value True or False. She will be able to access the resource if the verifier returns True otherwise she will not.

The steps are outlined here and the complete UML Sequence diagram is show in Figure 3:

- Each domain will have their own ABAC system consisting of the basic domain entities. Each domain entity is a separate decoupled microservice running on their own web server. The PEP only allows the names of the resources be available to the other domains. The contents are hidden as set by the administrator. The PEP also intercepts any request to the database which has the resources saved.
- The administrator write policies for the resources.
- Users from Domain 2 will be able to view the resource names and select a specific resource they want to view. This is provided by the PEP of domain 1. Domain 2 will also be able to send an access request to domain 1 to view the resource.
- The PEP in domain 1 would receive the request. It will be able to query the PAP to find the policy for the specific resource. After finding the specific policy, the PEP will turn the policy into a claim-predicate and send it to domain 2.
- The PEP of domain 2 would receive the claim-predicate. Then the PEP would need to know if domain 2 can satisfy it. So, the PEP would query the PIP to know if the attributes exist. This forms the basis of the decentralized PIP. Each domain will have their own PIP and will be able to query the attributes of the user each time any user of the domain wants to access resources in other domains.
- The PEP will receive the attributes. Using these attributes, it will generate a predicate statement that satisfies the claim-predicate. Then it will sign a message with the predicate.
- The PEP of domain 2 will send the signature to domain 1 for verification.
- The PEP of domain 1 receives the signature and verifies it to give a True or False response.

## C. Security Goal

The main idea of security goal is risk avoidance. For our model, we consider each domain to have an administrator who adds policies and resources. In a general sense, each user of the domain should be able to add resources and policies, however, for our implementation we only consider the administrator to be able to do that.

We assume the users of the domains do not collude with their attributes. This means multiple users should not be able to pool together their attributes and have access to only the attributes they possess. This is maintained by keeping the attributes stored in a database and using the PIP to query their attributes. Since the attribute authority in each domain could be a single point of attack as it is responsible for key generation, we consider certain security measures are taken to eliminate chances of attack. We assume that all the domain entities like PIP, PEP, PDP, PAP function properly.

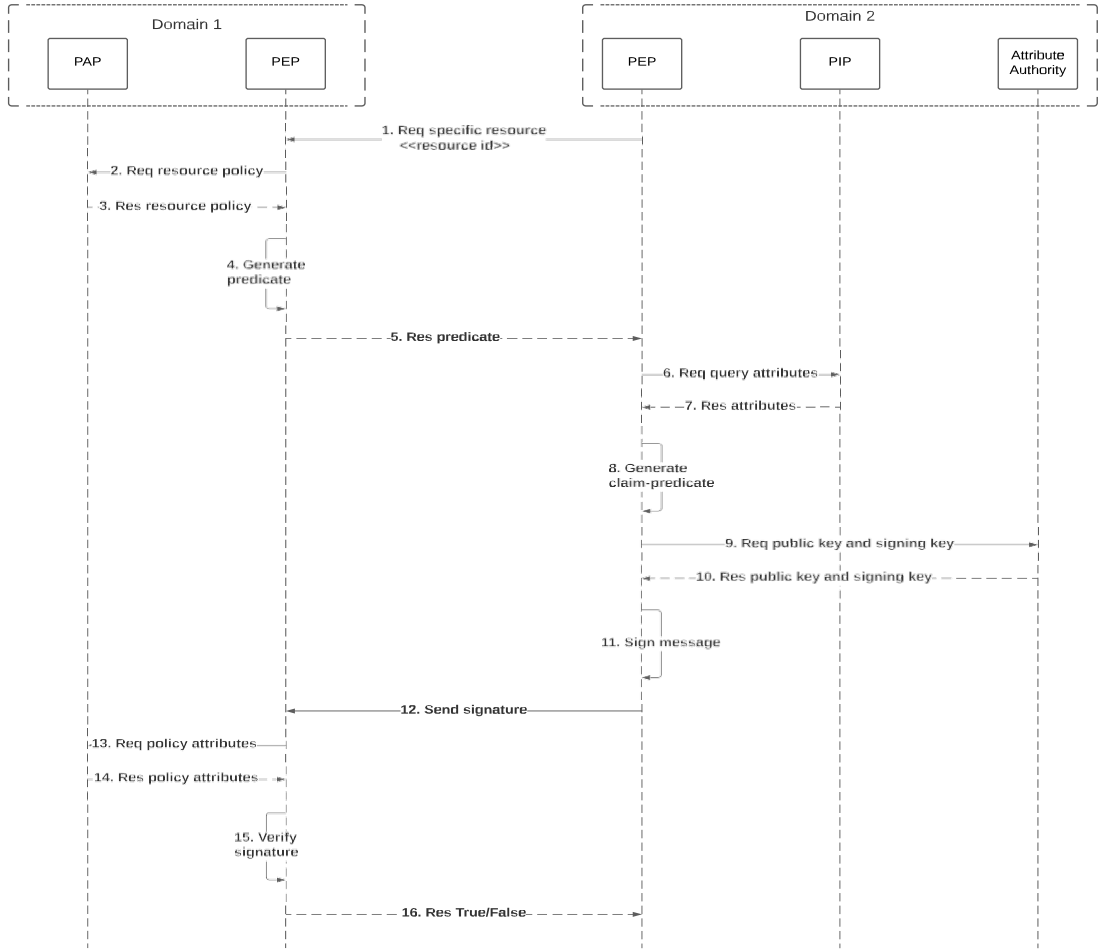

Fig. 3. The UML (Unified Modeling Language) Sequence Diagram of the Proposed Decentralized PIP model.

## V. IMPLEMENTATION

We used Django (v3.1), a python web framework to implement the web services [30]. Django supports authentication, development of web services and easily connects to databases which were essential for the development of our model.

Each domain entity of the ABAC system, PEP, PDP, PAP and PIP, was implemented as a microservice. Each domain entity runs as an autonomous process and communicate with other domain entities through web services provided by REST APIs.

In our model, the PAP provides the functionality for users to add resources and policies. The user interface of the PAP was implemented using the Django template engine and basic HTML, CSS. Jquery was used to make the pages more user friendly and intuitive [31]. Policies can be only set by the administrator of the domains. This means, that our model needed a user login-logout system. In-built User Authentication of the Django web framework was used to handle the administrator authentication [32].

We used an open source implementation of the ABS in our prototype [7] which uses an open-source framework called Charm-Crypto, which is written in Python, to implement the ABS algorithms [33]. Charm-Crypto is a framework for developing advanced cryptosystems. It is freely available to researchers and has been used in many researches [34].

## VI. EVALUATION

This section evaluates the implementation of the proposed Decentralized PIP model with the open source ABS. We are interested to see how quickly a user can access a resource. The evaluations show the usability of decentralized PIP. We consider the faster response time to be more usable as this means less time for signing and verification processes. Longer wait times are undesirable when trying to access resources. In typical use, the user will be presented with the contents of the resource that they requested. But there would be alot of processes going on in the back-end, which is hidden from the user. This evaluation section was conducted on a Linux machine with Core i7-10710U 12 core CPU running at 1.10GHz, 16GB RAM and Python 3.6.9 was used. We modeled five users with varying number of attributes.

In our model, domain 1 and domain 2 are two separate domains each running their own security configurations and web server. This simulates the multi-domain environment as they run in different ports. Each of the domains possess their own PIP to query attributes. This is to create decentralizedPIP environment which guarantees the users to not able to pool their attributes together. Each user is in charge of their own attributes. We imagine domain 1 possessing resources which means, domain 1 would be in charge of verifying signature to permit or deny access to their resource and for the administrator to enter policies. User from Domain 2 wouldsign messages based on claim-predicate. This signing process is very resource-intensive as the attribute authority of domain 2 would have to generate multiple keys and then sign a message using them. We list two methods to access resources:

- By generating new private and public keys and signature each time we want to access a resource.
- By generating public key, private key, signing key and

  signature once and storing according to their predicate in a database. Then querying the specific keys and signature when predicates match.

Here we evaluate our model to see which method is more efficient by finding a relation between the number of attributes and the time taken to get a response. The more efficient method would allow the access to the resource in less time which involves signing and verification.

In the graphs provided, the 'number of attributes' mean the attributes that are being used to generate the key and signature which is depicted in the x-axis of the graphs. The y-axis of the graphs show the time taken to get the response in seconds.

number of attributes, the more number of calculations need to be performed by the domain to verify.

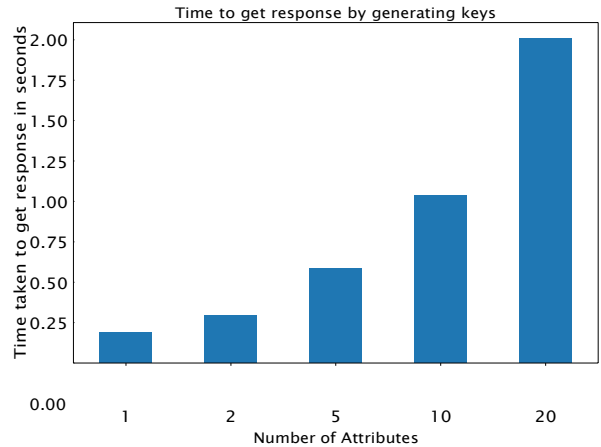

Fig. 5. Overall time taken to get response by generating keys.

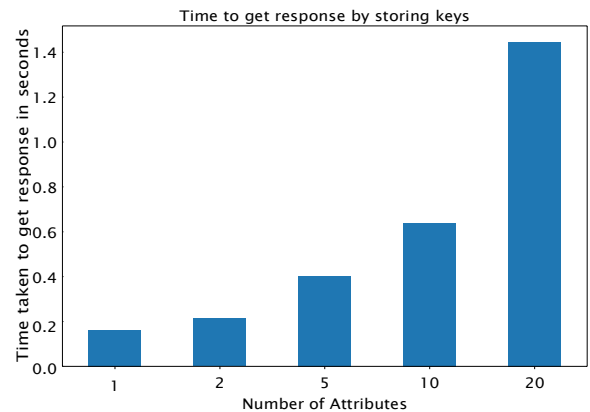

Fig. 6. Overall time taken to get response by storing keys.

The bar graphs in Figure 5 and 6 represent the time taken to get a response where the keys were generated each time and the time taken to get response where the keys were stored beforehand, respectively. This test was performed to see which method is quicker. They both show a linear relation between the time in seconds and the number of attributes. However, the time taken for each set of attribute values in stored key is less than the time taken by generating new keys each time. We can say that storing the keys before might be a viable option to achieve more efficiency.

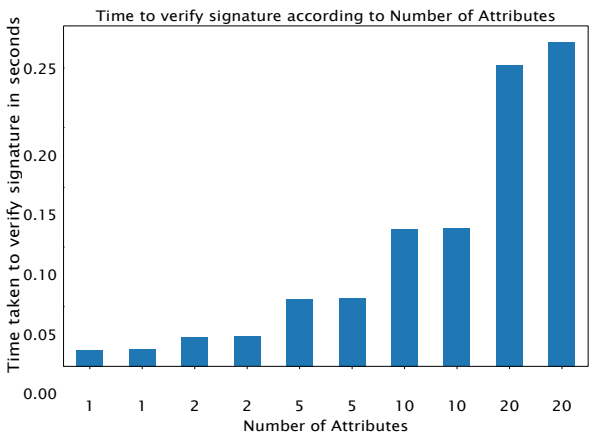

Fig. 4. Time to verify signature.

The bar graph in Figure 4 represents the time taken to verify signature based on the number of attributes. This graph has been generated based on the data from domain 1. Since we ran the prototype using two methods (key generation and key storage), there are duplicate number of attributes. The graph shows a linear relation between the time and the number of attributes. As the number of attributes increase, the time it takes to verify increases. This happens because the more the

The keys in our model were stored in a No-SQL database collection. Although it is fast, there might be performance bottlenecks in the case of querying through millions of records when the database gets heavy. This would add more time during the querying process which would affect the overall time. Storing keys in database might lead to attacks on the database where a third-party would be able to copy the keys and signature and might pose threats. Apart from this, we have not taken into account the geographic locations of the domains as it might add latency. The configurations of the servers might affect the speed of key generation too. As key generation requires mathematical computations. a server with low specifications would take longer to generate keys based on the same attributes whereas a server with high specification would be faster.

In conclusion, although generating keys each time has proved to be slower, it is the better option as it eliminates the need of storing keys in a database.

## VII. Conclusion and future work

In this paper, we proposed a Decentralized PIP model that makes use of attributes of users to grant or deny access to data. The proposed model provides more flexible and fine-grained AC than existing AC systems like RBAC and eliminates the risk of exposing private user information by implementing the concept of ABS. The implementation of a prototype demonstrated the feasibility of the proposed model.

Currently, our model only allows writing attribute names and values and does not support ranges of values when signingnew messages. In the future, we will add more functionalities in signing messages. We will also expand the implementation of the prototype to implement a more refined system anddeploy it on the cloud to make performance evaluations. Also we will make evaluations with other attribute based security schemes.